\documentstyle[preprint,aps,eqsecnum]{revtex} 

\input{psfig}
\def\be{\begin{equation}} 
\def\ee{\end{equation}}
\def\bea{\begin{eqnarray}} 
\def\eea{\end{eqnarray}}
\def\line{\hbox to \hsize}    
\def\frac #1#2{{#1\over #2}}

\def \v{{\bf v}}

\def\vev #1{{\langle #1\rangle}}
\def\1{\mbox{\bf 1}}

\begin{document}
\draft 

\title{
Acoustic Energy and Momentum in a Moving Medium
}

\author{ MICHAEL STONE}
\address{University of Illinois, Department of Physics\\ 1110 W. Green St.\\
Urbana, IL 61801 USA\\E-mail: m-stone5@uiuc.edu}   

\maketitle

\begin{abstract}

By exploiting the mathematical analogy between the propagation of
sound in a non-homogeneous potential flow and the propagation of a
scalar field in a background gravitational field, various wave
``energy'' and wave ``momentum'' conservation laws are established in
a systematic manner. In particular the acoustic energy conservation law
due to Blokhintsev appears as the result of the conservation of a
mixed co- and contravariant energy-momentum tensor, while the
exchange of relative energy between the wave and the mean flow mediated by the
radiation stress tensor,  first noted by Longuet-Higgins and Stewart
in the context of ocean waves, appears as the covariant conservation
of the doubly contravariant form of the same energy-momentum tensor.

\end{abstract}

\pacs{PACS numbers: 43.28.Py, 43.20.Wd, 43.25.Qp, 67.40.Mj   
  }

\section{Introduction}

Many discussions of the ``energy'' and ``momentum'' associated with
waves propagating  through moving fluids can be found in the
physics\cite{morse},
engineering\cite{blokhintsev,cantrell,ryshov,morfey,meyers86}, and
mathematical fluid mechanics literature
\cite{whitham62,whitham65,garrett,bretherton68,lighthill,andrews79a,andrews79b,longuet-higgins1,longuet-higgins2,meyers91}.
Various definitions are proposed, some of which lead to conserved
quantities, and some to quantities that are not conserved but
instead  exchanged between the wave and the mean flow.  In part the
multiplicity of definitions is due to difficulty  in deciding what
fraction of the energy or momentum of the system properly belongs to
the wave and what fraction should be associated with the moving 
medium. It is also often unclear how to divide equations expressing
conservation laws into terms relating to the conserved quantity, and terms
acting as sources for this quantity. Related to these primarily
cosmetic problems are more fundamental issues as to whether the
``energy'' or ``momentum'' under discussion is the true newtonian energy or
momentum, or instead pseudo-energy and pseudomomentum. Thus we have
the question ``what is the momentum of a sound wave'' raised  by Sir
Rudolf Peierls in his book {\it Surprises in Theoretical Physics\/}
\cite{peierls1}, and the salutary polemic ``{\it On the `Wave Momentum'
Myth\/}'' by M.~E.~McIntyre \cite{mcintyre}.

The most extensive  analyses of conserved wave properties have been
carried out by the fluid mechanics community
\cite{whitham62,whitham65,garrett,bretherton68,lighthill,andrews79a,andrews79b,longuet-higgins1,longuet-higgins2}.
Typically these papers adopt a lagrangian (following individual
particles in the flow) or mixed lagrangian-eulerian approach, as
opposed to the purely eulerian (describing the flow in terms of a
velocity field) approach which would be most familiar to a
physicist.  In addition a physicist reading this literature feels
the  lack of a general organising principle behind the definition and
derivation of the conservation laws. The present paper is intended to
remedy some of these problems --- at least  for the special case of
sound waves propagating through an irrotational homentropic  flow.
Although  rather a restricted class of motions, this is still one   of
considerable interest in condensed matter physics. It includes phonon
propagation  in a bose condensate, and therefore lies at the heart of
the two fluid model of superfluids.  By exploiting  W. Unruh's
ingenious identification \cite{unruh1,unruh2} of the wave equation
for sound waves in such a flow with the equation for a scalar field
propagating in a background gravitational field, I  extract the
conservation laws from the principle of general covariance. Deriving
the conservation laws in this way may seem like a case of taking a
sledge-hammer  to crack a nut, but the formalism is familar to most
physicists, automatic in application, and the
ambiguities in defining the conserved quantities turn out to lie in
the choice of whether to identify the energy-momentum tensor as
$T^{\mu\nu}$ or as $T^{\mu}_{\phantom{\mu}\nu}$. Also, when
quantities are not conserved, as is the case of the wave momentum in
a shear flow, their sources arise naturally from the connection
terms 
in the covariant derivative.

In section two I discuss the  action  describing  the irrotational
motion of an homentropic fluid. In section three I derive Unruh's equation from the
action principle.  In section four I explain why we often need
information beyond the solutions of linearized wave equation, and in
section five derive the conservation equations that follow from the
linearized equation.  Section six interprets these equations in
terms of the motion of phonons. The discussion  section repeats the
warnings from \cite{mcintyre} that the appearance  of   quantum
quasi-particles in any argument is a sure sign that the we are
considering pseudomomentum and not the actual momentum of the
system.

The work reported here was motivated by a desire to  understand better
the role of acoustic  radiation stress in  the two-fluid model of a superfluid. 
It may be relevent to the 
recent controversy\cite{wexler,wexlerthouless,sonin} over the {\it
Iordanskii force\/} acting on a vortex moving with respect to the
normal fluid component. The use of the Unruh formalism in this
context was suggested by Volovik\cite{volovik}.

\section{The Action Priciple}

The most straight forward way of deriving   conservation laws
starts from an action principle. Noether's theorem then provides us
with an explicit formula for a conserved quantity corresponding to
each symmetry of the action. In fluid mechanics, unfortunately, at
least when we restrict ourselves to an eulerian decription of the
flow field, action principles are in short supply.  Of course  there
must exist {\it some\/}  action principle  because ultimately the
fluid can be treated as a system of particles. A particle-based
action, however, requires a lagrangian description of the flow.  When
it is re-expressed in eulerian terms constraints appear, and these
limit its utility.
 
If  we restrict ourselves to flows that are both   irrotational and
homentropic --- the latter term meaning in practice that we assume the
pressure to be  a function  of the fluid density only ---  then the number
of degrees of freedom available to the fluid is  dramatically reduced.
In this case the  eulerian
equations of motion   {\it are\/} derivable from the
action\cite{schakel}
\be
S= \int d^4x\left\{ \rho \dot \phi +\frac 12 \rho (\nabla\phi)^2
+ u(\rho)\right\}.
\label{EQ:basic_action}
\ee
Here $\rho$ is the mass density, $\phi$ the velocity potential, and
the overdot denotes differentiation with respect to time. The
function $u$ may be identified with the internal energy density.  

Equating  to zero  the variation of  $S$ with respect to $\phi$
yields the continuity equation 
\be \dot \rho + \nabla\cdot (\rho \v)=0,
\label{EQ:continuity} \
\ee where  $\v\equiv\nabla \phi$.  Varying
$\rho$ gives Bernoulli's equation 
\be \dot\phi +\frac 12 \v^2 +
\mu(\rho)=0, 
\label{EQ:bernouilli} 
\ee where $\mu(\rho) = {d u}/{d \rho}$.
In  most applications $\mu$ would be identified with the specific
enthalpy.  For  a superfluid condensate  the entropy density, $s$, is
identically zero and  $\mu$ is the local chemical potential.

It is worth noting that our action could {\it not} have arisen from some
rewriting of the action for the motion of  a system of individual particles. We are
allowing variations of $\rho$ without requiring simultaneous
variations of $\phi$, and  such variations conjure new matter out of
nothing.

The gradient of the Bernoulli equation is 
Euler's equation of motion for the fluid. Combining this with the continuity equation
yields a  momentum conservation law
\be
\partial_t(\rho v_i)+\partial_j(\rho v_j v_i
) +\rho\partial_i \mu=0.
\label{EQ:momcons1} 
\ee
We simplify (\ref{EQ:momcons1}) by introducing the pressure, $P$, which is related to $\mu$ by   
$P(\rho)=\int\rho d\mu$. Then we can write 
\be
\partial_t(\rho v_i)+\partial_j \Pi_{ji}=0,
\ee
where   $\Pi_{ij}$ is
given by
\be 
\Pi_{ij}= \rho v_iv_j+\delta_{ij}P.
\ee
This is the usual form of the momentum flux tensor in  
fluid mechanics.

The relations $\mu= {d u}/{d \rho}$ and $\rho ={d P}/{d \mu}$ show that  
$P$ and $u$ are related by a Legendre transformation, 
$P= \rho\mu -u(\rho)$.
From this and the 
Bernoulli equation we see that
the pressure is equal to minus the action density,
\be
-P=   \rho \dot \phi +\frac 12 \rho (\nabla\phi)^2
+ u(\rho).
\ee
Consequently we can write 
\be
\Pi_{ij}= \rho\partial_i\phi\partial_j\phi-
\delta_{ij}\left\{ \rho \dot \phi +\frac 12 \rho (\nabla\phi)^2
+ u(\rho)\right\}.
\label{EQ:noether}
\ee 
This is the flux tensor that would appear were we  to
use  Noether's theorem to derive a law of momentum conservation
directly from the invariance of the action under the translation
$\phi({\bf r})\to \phi({\bf r} -{\bf a})$, $\rho({\bf r})\to
\rho({\bf r} -{\bf a})$. This is not a trivial point because there are
at  two similar, but distinct, notions of ``momentum''. True momentum
is associated with the symmetry of the action under a simultaneous
translation all the particles in the system. Its conservation
requires an absence of external forces. {\it
Pseudomomentum\/}\cite{peierls2} is the quantity that is preserved
when the action is left invariant when the {\it disturbance\/} in the
medium is relocated, but the reference position of each individual
particle is left unchanged. Conservation of pseudomomentum  requires
homogeneity of the medium rather than of space.  Replacing the field
$\phi({\bf r})$ by $\phi({\bf r} -{\bf a})$ would normally correspond
to the latter symmetry, but, because of the absence of explicit particles,
at this point in our discussion the two concepts coincide.

\section{The Unruh Metric}

We now  obtain the linearized wave  equation for  
the propagation of sound waves in a background
mean flow.  Let
\bea
\phi&=&\phi_{(0)} + \phi_{(1)}\nonumber\\
\rho&=&\rho_{(0)} + \rho_{(1)}. 
\eea
Here $\phi_{(0)}$ and $\rho_{(0)}$ define the mean flow. We assume  that they obey
the equations of motion. The quantities  
$\phi_{(1)}$ and $\rho_{(1)}$
represent small amplitude perturbations. 
Expanding $S$ to quadratic order in
these perturbations gives
\be
S=S_0+ \int d^4x\left\{\rho_{(1)} \dot \phi_{(1)}
+ \frac 12 \left(\frac {c^2}{\rho_{(0)}}\right) \rho_{(1)}^2
 +\frac 12 \rho_{(0)} (\nabla\phi_{(1)})^2
+ \rho_{(1)}\v\cdot\nabla \phi_{(1)} \right\}. 
\label{EQ:first_action}
\ee
Here $\v \equiv \v_{(0)}= \nabla\phi_{(0)}$.  The speed of sound, $c$, is defined by   
\be
\frac {c^2} {\rho_{(0)}}= \left.\frac {d \mu}{d \rho}\right|_{\rho_{(0)}}, 
\ee
or more familiarly
\be
c^2= \frac {d P}{d \rho}.
\ee
The terms linear in the perturbations  vanish because of
our assumption that the zeroth-order variables obey the equation of motion.

The equation of motion for  $\rho_{(1)}$ derived from
(\ref{EQ:first_action}) is 
\be
\rho_{(1)} = - \frac{\rho_{(0)}}{c^2}\{ \dot \phi_{(1)}+ \v\cdot\nabla
\phi_{(1)}\}.
\label{EQ:rhoeq}
\ee
In general we are not allowed to  substitute a consequence of  an
equation of motion back into the action integral. Here however, because  $\rho_{(1)}$ occurs
quadratically, we may use (\ref{EQ:rhoeq}) to eliminate it and   obtain an   
effective action for the  potential $\phi_{(1)}$ only 
\be
S_{(2)} = \int d^4x\left\{ \frac 12 \rho_{(0)} (\nabla\phi_{(1)})^2 - \frac {\rho_{(0)}} {2 c^2}
(\dot \phi_{(1)}+ \v\cdot\nabla \phi_{(1)})^2\right\}.
\label{EQ:-unruh_action}
\ee

The resultant equation of motion for $\phi_{(1)}$ is
\cite{unruh1,unruh2}
\be
\left(\frac{\partial}{\partial t}+\nabla\cdot\v\right)\frac
{\rho_{(0)}}{c^2}\left(\frac{\partial}{\partial t}+\v\cdot\nabla\right)\phi_{(1)}=
\nabla(\rho_{(0)}\nabla\phi_{(1)}).
\label{EQ:unruheq}
\ee
Note that in deriving this equation we have {\it not\/} assumed that the
background flow $\v$ is steady, only that it 
satisfies the equations of motion. Naturally, in order for our
waves to be distinguishable from the background flow, the latter should be
slowly changing  and have a  longer length scale than the wave motion.

Now (\ref{EQ:unruheq}) is equivalent to a perhaps more familiar
equation \cite{morse}
\be
\left(\frac{\partial}{\partial t}+\v\cdot\nabla\right)\frac
1{c^2}\left(\frac{\partial}{\partial t}+\v\cdot\nabla\right)\phi_{(1)}=
\frac 1 {\rho_{(0)}}\nabla(\rho_{(0)}\nabla\phi_{(1)}),
\label{EQ:usual}
\ee
as can be seen by using the mass conservation equation $\partial_t\rho_{(0)} +\nabla \cdot\rho_{(0)}\v=0$, 
but the form (\ref{EQ:unruheq}) 
has the advantage that it can be written as\footnote{I use the convention that greek letters 
run over four space-time 
indices  $0,1,2,3$ with $0\equiv t$, while roman indices refer to the three space
components.}
\be
\frac 1{\sqrt{-g}} \partial_\mu {\sqrt{-g}}g^{\mu\nu}\partial_\nu
\phi_{(1)}=0,
\label{EQ:scalar_eq}
\ee
where 
\be
\sqrt{-g}g^{\mu\nu} = \frac {\rho_{(0)}}{c^2}\left(\matrix{ 1, & \v^T \cr
                                                            \v,& \v\v^T - c^2{\bf
1}\cr}\right).
\label{EQ:unruh_metric_up}
\ee  
This is perhaps most easily seen by observing that   
the action (\ref{EQ:-unruh_action}) is equal to $-S$ where 
\be 
S= \int d^4x\,\frac 12  \sqrt{-g} g^{\mu\nu} \partial_\mu \phi_{(1)} \partial_\nu
\phi_{(1)}=  \int d^4x\, \sqrt{-g}L.
\ee

Equation  (\ref{EQ:scalar_eq})   has the same form as  that of a
scalar wave propagating  in a  gravitational field with
Riemann metric $g_{\mu\nu}$. The idea of writing the  acoustic wave
equation in this way, as well as the general relativity analogy, is
due to Unruh \cite{unruh1,unruh2}.  I will therefore refer to
$g_{\mu\nu}$ as the Unruh metric.

The $4$-volume measure   $\sqrt{-g}$
is equal to $ \rho_{(0)}^2/c$, and
the covariant components of the metric are
\be
g_{\mu\nu}= \frac {\rho_{(0)}}{c}\left(\matrix{ c^2-v^2, & \v^T \cr
                                                            \v,& -{\bf
1}\cr}\right).
\label{EQ:unruh_metric_down}
\ee
The associated space-time interval is therefore
\be
ds^2= \frac {\rho_{(0)}} c \left\{c^2dt^2
-\delta_{ij}(dx^i-v^idt)(dx^j-v^jdt)\right\}.
\ee
Metrics of this form, although without the overall conformal factor
${\rho_{(0)}}/{c}$, appear in the  Arnowitt-Deser-Misner (ADM)
formalism of general relativity\cite{ADM}.  There $c$ and $-v^i$
are refered to as  the lapse function and shift vector repectively.
They serve to glue successive three-dimensional time slices
together to form a four dimensional space-time\cite{MTW}. In our
present case, provided  ${\rho_{(0)}}/{c}$ can be regarded as a constant, 
each $3$-space is ordinary flat ${\bf R^3}$ equipped with
the rectangular cartesian metric $g^{(space)}_{ij}=\delta_{ij}$ ---
but the resultant {\it space-time\/} is in general curved, the
curvature depending on the degree of inhomogeneity of the mean flow
$\v$.

In the geometric acoustics limit sound  will travel along the null
geodesics defined by $g_{\mu\nu}$. Even in the presence of spatially
varying $\rho_{(0)}$ we would expect the ray paths to depend only on
the local values of $c$ and $\v$, so it  is perhaps a bit
surprising to see the density entering the expression for
the Unruh metric. However an overall conformal factor does not affect  {\it null\/}
geodesics, and thus variations in $\rho_{(0)}$ do not influence the ray
tracing. For steady flow, and in the case that
only $\v$ is varying, it is  shown in the appendix that the null
geodesics coincide with the ray paths obtained by applying Hamilton's
equations for rays
\be 
\dot x^i = \frac{\partial \omega}{\partial k_i},
\qquad \dot k_i=- \frac{\partial \omega}{\partial
x^i},
\ee
to the appropriate Doppler shifted frequency
\be
\omega({\bf x}, {\bf k})= c|{\bf k}| + \v\cdot {\bf k}.
\ee 

When $\v$ is  in the $x$
direction only, we can also rewrite $ds^2$ as
\be
ds^2=\frac {\rho_{(0)}} {c} \left\{-\left(dx-(v+c)dt\right)\left(dx-(v-c)dt\right)
-dy^2-dz^2\right\}.
\label{EQ:unruh_simplified}
\ee
This shows  that the $x-t$ plane  null geodesics coincide with the expected
characteristics of the wave equation in the background flow.

\section{Momentum Flux}

The fluid in a sound wave has average velocity zero, but since the
fluid is compressed  in the half cycle when it is moving in the
direction of  propagation and rarefied when it is moving backwards
there is a net mass current (and hence a momentum density) which is
of second order in the sound wave amplitude $a_0$.  
This becomes clearer if one solves the equation
\be
\frac{dx}{dt}=v(x)=a_0\cos(kx-\omega t)
\ee
for the trajectory $x(t)$ of a fluid particle. This equation is non-linear
($x$ appears inside the cosine), and solving perturbatively one
finds a secular drift at second order in $a_0$. 
\be
x(t)= x(0)+ \hbox{\rm oscillations} + \frac {1}{2}a_0^2 \left(\frac
k\omega\right) t +\cdots .
\ee
Although the
time average of the eulerian fluid velocity, $v$,  is zero, the time
average of the {\it lagrangian\/} velocity, $v_L=\dot x$, is not.   
The difference beweeen the two
average velocities is the  {\it Stokes drift\/}.  The Stokes drift is
$O(a_0^2)$ while the wave equation is accurate only to $O(a_0)$, so
care is necessary before using its solutions to evaluate the mass
current.  Similar problems occur in defining the energy density and
energy and momentum fluxes which also require second order accuracy.

We can expand the velocity field as   
\be
\v= \v+\v_{(1)}+\v_{(2)}+\cdots,
\ee
where the second-order correction $\v_{(2)}$ arises as as consequence
of the  nonlinearities in the equations of motion.  This correction
will possess both oscillating and steady components. The oscillatory components arise 
because a
strictly harmonic  wave with frequency $\omega_0$ will gradually
develop  higher frequency components due to  the progressive
distortion of the wave as it propagates. (A plane wave eventually
degenerates into a sequence of  shocks.). These distortions  are
usually not significant in considerations of energy and momentum
balance.   The steady terms, however, represent $O(a_0^2)$
alterations to the  mean flow caused  by the sound waves, and these
often  possess  energy and momentum  comparable to that of
the  sound field.

Even if we temporarily ignore these effects and retain only
$\v_{(1)}$ as determined from the linearized wave equation, the
density and pressure will still have expansions
\bea
\rho&=&\rho_{(0)} + \rho_{(1)}+\rho_{(2)}+\cdots \nonumber\\
P&=&P_{(0)} + P_{(1)}+P_{(2)}+\cdots.
\eea
As before,  the grading $(n)$  refers to the number of powers of the sound wave
amplitude in an expression. The small parameter in these expansions is
the Mach number given by a typical value of $v_{(1)}$ divided by the
local speed of sound.

Consider for example the momentum density $\rho\v$ and the momentum flux
\be
\Pi_{ij} =\rho v_iv_j +\delta_{ij} P.
\ee
It is reasonable to define the  momentum density and the momentum flux
tensor associated with the sound field
as the second order terms
\be
{\bf j}^{\rm (phonon)}=\vev{\rho\v}= \vev{\rho_{(1)}\v_{(1)}} + \v\vev{\rho_{(2)}},
\ee
and 
\be
\Pi_{ij}^{\rm (phonon)}= \rho_{(0)}\vev{v_{(1)i}v_{(1)j}}
+v_{i}\vev{\rho_{(1)}v_{(1)j}}+v_{j}\vev{\rho_{(1)}v_{(1)i}}
+\delta_{ij}\vev{P_{(2)}} + v_{i}v_{j}\vev{\rho_{(2)}}.
\ee
(The angular brackets indicate that we should take a time average over a sound wave period.
There is no need to consider  terms first order in the amplitude because these  average to zero.)  
We see that to compute them we need  to consider the second order
contributions to both $P$ and $\rho$. 

We can compute $P_{(2)}$ in terms of first order quantities from
\be
\Delta P= \frac{d P}{d\mu}\Delta \mu + \frac 12  \frac{d^2
P}{d\mu^2}(\Delta\mu)^2 +O((\Delta\mu)^3)
\ee
and Bernoulli's equation in the form
\be
\Delta \mu = - \dot \phi_{(1)} - \frac 12 (\nabla \phi_{(1)})^2 - \v\cdot \nabla \phi_{(1)}, 
\ee
together with
\be
\frac{d P}{d\mu}= \rho,\qquad  \frac{d^2
P}{d\mu^2}=\frac{d \rho}{d\mu} =
\frac{\rho}{c^2}.
\ee
Expanding out and grouping terms of appropriate orders gives
\be
P_{(1)}= -\rho_{(0)} (\dot\phi_{(1)} + \v\cdot\nabla \phi_{(1)}) = c^2\rho_{(1)},
\label{EQ:P1}
\ee
which we already knew, and
\be
P_{(2)} =-\rho_{(0)}\frac 12 (\nabla \phi_{(1)})^2 +\frac 12 \frac{\rho_{(0)}}{c^2}(\dot\phi_{(1)} + \v\cdot\nabla
\phi_{(1)})^2.
\label{EQ:P2}
\ee
We see that $P_{(2)}=\sqrt{-g}L$ where $L$ is the  Lagrangian density for our sound wave equation.

To extract $\rho_{(2)}$ in this manner we need more information
about  the equation of state of the fluid than is used in the
linearized theory.  This information is  most conveniently parameterized by 
the logarithmic derivative of the speed of sound with
pressure (a fluid-state physics analogue of the Gr{\"u}neisen parameter). 
Using this  together with the previous results for $P_{(2)}$, we
find 
that
\be
\rho_{(2)}= \frac 1{c^2} P_{(2)} - \frac
1{\rho_{(0)}} \rho_{(1)}^2 \left.\frac{d\ln c}{d\ln \rho}\right|_{\rho_{(0)}}.
\label{EQ:true_rho2}
\ee

\section{Conservation Laws}

While  we cannot compute the ``true'' energy and momentum densities
and fluxes without including non-linear corrections to the motion, it
is often more useful find closely related  quantities
whose conservation laws are a consequence of the linearized wave equation,
and which therefore provide information about the solutions of this
equation. Our ``general relativistic'' formalism provides a sytematic
way  of finding such conserved quantities.  It is well known
\cite{weinberg} that any action $S$ automatically provides us with a
covariantly conserved and  symmetric energy-momentum tensor
$T_{\mu\nu}$ defined by
\be
T_{\mu\nu}= \frac 2{\sqrt{-g}}\frac{\delta S}{\delta g^{\mu\nu}}.
\ee
The functional derivative is here defined by
\be 
\delta S= \int d^4x \, \sqrt{-g}\frac{\delta S}{\delta g^{\mu\nu}}
\delta g^{\mu\nu}.
\ee 

It follows from  the equations of motion derived from $S$ that
\be
D_\mu T^{\mu\nu}=0,
\ee
where  $D_\mu$ is the covariant
derivative. For example 
\be
D_\alpha A^{\mu\nu}_{\phantom{\mu\nu}\sigma}= \partial_\alpha A^{\mu\nu}_{\phantom{\mu\nu}\sigma}+
\Gamma^\mu_{\alpha\gamma} A^{\gamma\nu}_{\phantom{\mu\nu}\sigma}+ \Gamma^\nu_{\alpha\gamma}
A^{\mu\gamma}_{\phantom{\mu\nu}\sigma}- \Gamma^\gamma_{\alpha\sigma}
 A^{\mu\nu}_{\phantom{\mu\nu}\gamma}.
\ee
The   $\Gamma^\alpha_{\beta\gamma}$ are  
the components of the Levi-Civita connection 
compatable with the Unruh metric, {\it viz.\/}
 \be
\Gamma^\alpha_{\beta\gamma} = g^{\alpha\rho}[\beta\gamma,\rho],
\ee
where
\be
[\beta\gamma,\rho]= \frac12 \left( \frac{\partial
g_{\gamma\rho}}{\partial x^\beta} + \frac{\partial
g_{\beta\rho}}{\partial x^\gamma} - \frac{\partial
g_{\beta\gamma}}{\partial x^\rho}\right).
\label{EQ:christoffel_def} 
\ee

For our scalar field 
\be
T^{\mu\nu}= \partial^\mu \phi_{(1)}\partial^\nu \phi_{(1)}-
g^{\mu\nu}\left(\frac 12 g^{\alpha\beta}
\partial_\alpha\phi_{(1)}\partial_\beta\phi_{(1)}\right).
\label{EQ:Tmunu}
\ee
The derivatives with raised indices in (\ref{EQ:Tmunu}) are
defined by
\be
\partial^0 \phi_{(1)} = g^{0\mu}\partial_\mu\phi_{(1)}= \frac
1{\rho_{(0)}c} (\dot\phi_{(1)} + \v\cdot\nabla \phi_{(1)}),
\ee
and
\be
\partial^i \phi_{(1)}=g^{i\mu}\partial_\mu\phi_{(1)}=   
\frac
1{\rho_{(0)}c} \left( v_{i}(\dot\phi_{(1)} + \v\cdot\nabla \phi_{(1)})
   -c^2\partial_i \phi_{(1)}\right).
\ee

Thus 
\bea
T^{00}&=& \frac{1}{\rho_{(0)}^3}\left(\rho_{(0)}\frac 12 (\nabla \phi_{(1)})^2 +\frac 12 
\frac{\rho_{(0)}}{c^2}(\dot\phi_{(1)} + \v\cdot\nabla
\phi_{(1)})^2\right)\nonumber\\ 
&=& \frac{c^2}{\rho_{(0)}^3} \left(\frac {W_r}{c^2}\right) \nonumber\\
&=&\frac{c^2}{\rho_{(0)}^3}\tilde\rho_{(2)}.
\label{EQ:TOO}
\eea 
The last two equalities serve as a definition of  $W_{r}$ and  $\tilde\rho_{(2)}$.
The quantity  $W_{r}$ is often   decribed as the acoustic energy density relative to
the frame moving
with the local fluid velocity\cite{lighthill}. Because its
conservation law will depend on  the steadiness of the flow rather than the absence of 
time-dependent external forces, it is more correctly a 
pseudo-energy density.

Using (\ref{EQ:P1}), and (\ref{EQ:P2}) in the form
\be
\frac 12 g^{\alpha\beta}
\partial_\alpha\phi_{(1)}\partial_\beta\phi_{(1)}= \frac
{c}{\rho_{(0)}^2}P_{(2)},
\ee
we can express the other components  of (\ref{EQ:Tmunu}) in terms of
physical quantities.  We  find that
\bea
T^{i0}=T^{0i}&=& \frac{c^2}{\rho_{(0)}^3}\left(\frac 1{c^2} (P_{(1)}v_{(1)i} +
v_{i}W_r)\right)\cr
&=&\frac{c^2}{\rho_{(0)}^3}\left( \rho_{(1)}v_{(1)i}+
v_{i}
\tilde\rho_{(2)}\right).
\eea
The first  line in this expression shows that, up to an overall factor, $T^{i0}$ is the energy
flux -- the first term being the rate of working  by a fluid
element on its neigbour, and the second the advected energy.
The second  line is written so as to suggest the usual relativistic 
identification  of 
(energy-flux)$/c^2$ with  the density of momentum. 
This  interpretation, however, requires that
$\tilde\rho_{(2)}$ be  the  second order correction to the density, which it is not.  

Similarly
\be
T^{ij}=\frac{c^2}{\rho_{(0)}^3}\left(\rho_{(0)}v_{(1)i}v_{(1)j}
+v_{i}\rho_{(1)}v_{(1)j}+v_{j}\rho_{(1)}v_{(1)i}
+\delta_{ij}P_{(2)} + v_{i}v_{j}\tilde\rho_{(2)}\right).
\ee
We again see that if we identify $\tilde\rho_{(2)}$ with $\rho_{(2)}$
then $T^{ij}$ has  the exactly the  form as  we expect for the
second order momentum flux tensor.

The reason why the  linear theory  makes  the erroneous  
identification of  $\rho_{(2)} $ with $\tilde\rho_{(2)}$
is best  seen if we
set $\v=const.$ Then the equation 
\be
\partial_t T^{00}+ \partial_i T^{i0}=0,
\ee
holds. This reads
\be
\frac{c^2}{\rho_{(0)}^3}\left(\partial_t \tilde\rho_{(2)}
+\partial_i(\rho_{(1)}v_{(1)i}+ v_{i}\tilde\rho_{(2)})\right)=0,
\ee
and  looks very much like   the second order continuity equation 
\be 
\partial_t \rho_{(2)}
+\partial_i(v_{(2)}\rho_{(0)}+\rho_{(1)}v_{(1)i}+ v_{i}\rho_{(2)})=0,
\label{EQ:2nd_continuity}
\ee
once we  ignore $\v_{(2)}$.
When we go beyond the linear  theory (\ref{EQ:2nd_continuity}) provides a source or sink term
in the mass conservation  equation for  $\vev{\v_{(2)}}$ \cite{lighthill_streaming}, and is 
not    an equation determining
$\rho_{(2)}$.

We can also write the mixed co- and contra-variant components of the energy 
momentum tensor  $T^{\mu}_{\phantom {\mu}\nu}=
T^{\mu\lambda}g_{\lambda\nu}$
in terms of physical quantities. This mixed tensor  turns out to be  
more useful than the doubly contravariant tensor.  Because 
we no longer  enforce a symmetry  between the indices $\mu$ and $\nu$, 
the quantity $W_r$ is no longer
required to perform double duty as both an energy and a density. We find   
\bea
T^{0}_{\phantom {0}0} &=&\frac{c}{\rho_{(0)}^2} \left( W_r+
\rho_{(1)}\v_{(1)}\cdot \v\right)\cr
T^{i}_{\phantom {i}0}&=& \frac{c}{\rho_{(0)}^2}
\left( \frac {P_{(1)}}{\rho_{(0)}} + 
\v\cdot \v_{(1)}\right)(\rho_{(0)}v_{(1)i}+\rho_{(1)}v_{(0)i}),
\eea 
and 
\bea
T^{0}_{\phantom {0}i}&=& -\frac{c}{\rho_{(0)}^2}\rho_{(1)}\v_{(1)i}\cr
T^{i}_{\phantom {i}j}&=& -\frac{c}{\rho_{(0)}^2}\left(\rho_{(0)}v_{(1)i}v_{(1)j}
+ v_i\rho_{(1)}v_{(1)j} + \delta_{ij}P_{(2)}\right). 
\eea
We see that $\tilde \rho_{(2)}$ does not appear here, and all these
terms may be identified with physical quantities which are  reliably computed 
from solutions of the linearized wave equation.

The covariant conservation law can be written as either
$D_\mu T^{\mu\nu}=0$ or as $ D_\mu T^{\mu}_{\phantom {\mu}\nu}=0$.
The two equations are consistent with each other because the
covariant derivative is  defined so that$ D_\lambda g_{\mu\nu}=
g_{\mu\nu} D_\lambda$. To extract the   physical meaning of these equations
we need to evaluate the the connection forms $\Gamma^\mu_{\nu\lambda}$.

In what follows I will consider only a steady  background flow, and further one 
for which $\rho_0$, $c$, and hence 
$\sqrt{-g}=\rho_{(0)}^2/c$ can be treated as constant.
To increase the readabilty of some  expressions I will also
choose units so that $\rho_0$ and $c$ become unity and no longer
appear as overall factors in the metric or the four dimensional
energy-momentum tensors. I will however reintroduce them when they
are required for  dimensional correctness in expressions such as
$\rho_{(0)}\v_{(1)}$ or $W_r/c^2$.

From the Unruh metric we find
\bea
[ij,k] &=& 0\cr
[ij,0] &=& \frac 12 (\partial_i v_{j} + \partial_j v_{i})\cr
[i0,j] &=& \frac 12 (\partial_i v_j-\partial_j v_i)\cr
[0i,0] &=& [i0,0]= -\frac 12 \partial_i |v|^2\cr
[00,i] &=&  \frac 12 \partial_i |v|^2\cr
[00,0] &=& 0.
\eea
I have retained the expression $\frac 12 (\partial_i v_j-\partial_j
v_i)$ in $[i0,j]$, since it is possible that 
our wave equation has greater generality than its derivation.

We therefore find 
\bea
\Gamma^0_{00}&=& \frac 12 (\v \cdot \nabla) |v|^2 \cr
\Gamma^0_{i0}&=& -\frac 12 \partial_i\, |v|^2 +\frac 12 v_j(\partial_i v_j-\partial_j v_i)\cr
\Gamma^i_{00}&=& \frac 12 v_{i} (\v \cdot \nabla) |v|^2  - \frac 12 \partial_i\, |v|^2  \cr
\Gamma^0_{ij}&=&  \frac 12 (\partial_i v_{j} + \partial_j v_{i}) \cr
\Gamma^i_{j0}&=&  -\frac 12 v_{i}\partial_j\, |v|^2  
+\frac 12 (\partial_j v_k - \partial_k v_j)(v_kv_i-c^2\delta_{ik}) \cr
\Gamma^i_{jk}&=&  \frac 12  v_{i}(\partial_j v_{k} + \partial_k
v_{j}).
\eea

From (\ref{EQ:christoffel_def})  we have  
\be
\Gamma^\mu_{\mu\beta}= \frac{1}{\sqrt{-g}} \frac{\partial\sqrt{-g}}{\partial
x^\beta},
\ee
so, with $\sqrt{-g}=const.$, the trace $\Gamma^\mu_{\mu\beta}$ is
zero. One may verify that the above expressions for
$\Gamma^\mu_{\nu\lambda}$ obey this identity.

We now  evaluate 
\bea
D_\mu T^{\mu 0} &=&  \partial_\mu  T^{\mu
0} + \Gamma^\mu_{\mu\gamma} T^{\gamma 0}+\Gamma^0_{\mu\nu} T^{\mu\nu}\cr
   &=& \partial_\mu T^{\mu
0} + \Gamma^0_{\mu\nu} T^{\mu\nu}.
\eea
After a little algebra we find 
\be
 \Gamma^0_{\mu\nu} T^{\mu\nu} = \frac 12 (\partial_i v_j +
\partial_j v_i)(\rho_{(0)} v_{(1)i} v_{(1)j}+ \delta_{ij}P_{(2)}). 
\ee
Note the  non-appearence of  $\rho_{(1)}$
and $\tilde\rho_{(2)}$ in the final expression --- even though both quantities appear in
$ T^{\mu\nu}$.

The conservation law therefore becomes 
\be
\partial_t W_r +\partial_i (P_{(1)}v_{(1)i}+v_iW_r) + \frac 12 \Sigma_{ij} (\partial_i v_j +
\partial_j v_i)=0,
\ee
where
\be
\Sigma_{ij}= \rho_{(0)} v_{(1)i} v_{(1)j}+ \delta_{ij}P_{(2)}.
\ee
This is  an example of the general form of  energy law derived by
Longuet-Higgins and Stuart, originally in the context of ocean
waves\cite{longuet-higgins1,longuet-higgins2}.  (See also
\cite{ryshov} for a slightly earlier, but less general, case.) The
relative energy density, $W_r\equiv T^{00}$, is {\it not\/} conserved.
Instead an observer moving with the fluid  sees the waves acquiring
energy from the mean flow at a rate given by the product of a
radiation stress $\Sigma_{ij}$ with the mean-flow rate of strain.
Such non-conservation is not surprising. Seen from the viewpoint of the
moving frame the flow is no longer steady, while (pseudo) energy
conservation requires a time-independent medium.

 Notice that, since we are assuming that $\rho_{(0)}$ is a
constant, we should  for consistency require $\nabla \cdot \v=0$.
Thus the {\it isotropic\/} part of the radiation stress (the part
$\propto \delta_{ij}$) does no work. This is fortunate because the
non-linear theory shows that the isotropic radiation stress contains a
part dependent on $ {\partial \ln c}/{\partial \ln \rho}$ which is
missed by the  linear approximation. (see
however,\cite{bretherton_lec})

We  now examine  the energy conservation law coming from the zeroth component of the mixed  
energy-momentum tensor. We need  
\bea
D_\mu T^{\mu}_{\phantom{\mu}0} &=& 
 \partial_\mu T^{\mu}_{\phantom{\mu} 0}
-  \Gamma^\rho_{\mu 0} T^{\mu}_{\phantom{\mu}\rho}\cr
 &=&  \partial_\mu T^{\mu}_{\phantom{\mu} 0} - [\mu 0,\rho]
T^{\mu\rho}\cr
 &=&  \partial_\mu T^{\mu}_{\phantom{\mu} 0} -[i 0,0]
T^{i0}- [0 0,i]
T^{0i}- [i0,j]T^{ij}.
\eea
We now observe that $T^{i0}=T^{0i}$ while $[0 0,i]=- [i 0,0]$, and 
that $[i0,j]=- [j0,i]$, while $T^{ij}=T^{ji}$.
Thus the  connection  contribution vanishes. 
This form of the energy conservation law
is therefore
\be
\partial_t\left( W_r+
\rho_{(1)}\v_{(1)}\cdot \v\right)+ \partial_i\left( (\frac {P_{(1)}}{\rho_{(0)}} + 
\v\cdot \v_{(1)})(\rho_{(0)}v_{(1)i}+\rho_{(1)}v_{(0)i})\right)=0.
\label{EQ:blokhintsev}
\ee
Here we see that the combination $ W_r+ \rho_{(1)}\v_{(1)}\cdot \v$
{\it does \/} correspond to a conserved energy.  This 
conservation law was originally derived  by
Blokhintsev\cite{blokhintsev}  for slowly varying flows, and more
generally by Cantrell and Hart\cite{cantrell} in their study of the
acoustic stability of rocket engines. See also reference \cite{morfey},
and  \cite{andrews79b} eq. (5.18). 

Now we turn to the equation for momentum conservation.
Working similarly to the energy law we find 
\bea
D_\mu T^{\mu}_{\phantom{\mu}j}&=& 
\partial_\mu T^{\mu}_{\phantom{\mu}j} -[\mu j,\rho] T^{\mu\rho}\cr
&=& \partial_\mu T^{\mu}_{\phantom{\mu}j} - [0j,0] T^{00} - [ij,0] T^{i0}-[0j,i]T^{0i}\cr
 &=& \partial_\mu  T^{\mu}_{\phantom{\mu}j} 
 - \rho_{(1)}v_{(1)i}  \partial_j v_i.
\label{EQ:momnoncon}
\eea
Again notice the cancellation   of the terms containing $\tilde\rho_{(2)}$.  

The covariant conservation equation  $D_\mu T^{\mu}_{\phantom{\mu}j}=0$ therefore reads 
\be
\partial_t \rho_{(1)}v_{(1)j} + \partial_i \left(\rho_{(0)}v_{(1)i}v_{(1)j}
+ v_i\rho_{(1)}v_{(1)j} + \delta_{ij}P_{(2)}\right)
+ \rho_{(1)}v_{(1)i}  \partial_j v_i=0.
\ee
The connection terms  have provided a source term for the momentum density.
Thus, in an inhomogeneous flow field,  momentum is exchanged beween the
waves and the mean flow.

\section{Phonons and Conservation of Wave Action}

If the mean flow changes only slowly over many wavelengths,
the sound 
field can  locally be approximated by  a plane wave  
\be
\phi(x,t)= a_0 \cos({\bf k}\cdot {\bf x} -\omega t).
\ee
The frequency $\omega$ and the wave-vector $\bf k$ are here related by the 
Doppler-shifted dispersion relation 
$
\omega= \omega_r + {\bf k}\cdot{\v},
$
where the relative frequency,  $\omega_r=c|k| $, is that measured in the frame moving with the
fluid.  A packet of such waves 
moves at the group velocity 
\be
{\bf U} =\dot {\bf x}=  c \frac{{\bf k}}{|k|} +\v.
\ee
As the wave progresses through regions of varying $\v$, the parameters 
$\bf k$ and  $a_0$ will slowly evolve. The change in  ${\bf k}$ 
is given by the ray tracing formula
(\ref{EQ:kdot})
\be
\frac{d k_j}{dt} =- k_i \frac{\partial v_j}{\partial
x^j},
\ee
where the  time derivative is taken along the ray
\be
 \frac{d }{dt} = \frac{\partial}{\partial t} + {\bf U}\cdot \nabla.
\ee   
The evolution of the amplitude $a_0$ is linked with that of the energy
density, $W_r$ through
\be
\vev{W_r}= \frac 12 a_0^2 \rho_{(0)}
\frac{\omega_r^2}{ c^2}.
\ee

Now 
for a  homogeneous stationary fluid we would expect our    
macroscopic plane wave to correspond to  a  quantum coherent state whose energy 
is given in terms of  the (quantum) average phonon density $\bar N$ as
\be
E_{tot}= (\hbox{\rm Volume}) \vev{ W_r} =(\hbox{\rm Volume}) \bar N\hbar \omega_r.    
\ee
Since it is a density of ``particles'', 
$\bar N$ should remain the same when viewed from any frame, consequently   the relation
\be
\bar N\hbar= \frac{\vev{W_r}}{\omega_r}
\ee
should hold true generally. In classical fluid mechanics the quantity   
$\vev{W_r}/\omega_r$ is called the {\it wave action\/}\cite{garrett,lighthill,andrews79b}.

The time averages of  other components of the energy momentum 
tensor may be also expressed in terms of $\bar N$.
For the mixed tensor we have
\bea
\vev{T^0_{\phantom{0}0}}&=&\vev{W_r + \v\cdot \rho_{(1)} \v_{(1)}}  = \bar N\hbar
\omega\cr
\vev{T^i_{\phantom{i}0}} &=& \vev{( \frac {P_{(1)}}{\rho_{(0)}} +\v\cdot
\v_{(1)})( \rho_{(0)}v_{(1)i}+\rho_{(1)}v_i)} =\bar N\hbar \omega
U_i\cr
\vev{-T^0_{\phantom{0}i}}&=&\vev{\rho_{(1)} v_{(1)i} }=  \bar N\hbar 
k_i\cr
  \vev{-T^i_{\phantom{i}j}} &=&  \vev{ \rho_{(0)} v_{(1)i}v_{(1)j} 
+ v_i  \rho_{(1)} v_{(1)j} + \delta_{ij} P_{(2)}} =  \bar N\hbar k_j  U_i.
\eea
The last result uses the fact that $\vev{P_{(2)}}=0$ for a plane progressive wave.

Inserting these approximate expressions for the time averages into the
Blokhintsev energy  conservation law (\ref{EQ:blokhintsev}) we find that
\be
\frac{\partial \bar N\hbar \omega}{\partial t} + \nabla\cdot 
(\bar N\hbar \omega {\bf U})=0.
\ee
We can write this as
\be
\bar N\hbar\left(\frac{\partial  \omega}{\partial t}+ {\bf U}\cdot \nabla
\omega\right)
+ \hbar\omega\left( \frac{\partial \bar N }{\partial t} + \nabla\cdot 
(\bar N {\bf U})\right)=0.
\ee  
The first term is equal to  ${d\omega}/{dt}$ along the rays and  
vanishes for a steady mean flow as a consequence of the hamiltonian nature
of the ray tracing equations. The second term
must therefore also vanish. This represents the conservation of
phonons, or in classical language, the conservation of
wave-action.

In a similar manner the time average of (\ref{EQ:momnoncon}) may be written
\bea
0 &=& \frac{\partial \bar N k_j}{\partial t} + \nabla\cdot 
(\bar N k_j {\bf U}) + \bar N k_i \frac{\partial v_i}{\partial
x^j}\cr
 &=& \bar N\left(\frac{\partial k_j}{\partial t}+ {\bf U}\cdot \nabla
k_j+ k_i\frac{\partial v_i}{\partial
x^j} \right)
+  k_j\left( \frac{\partial \bar N }{\partial t} + \nabla\cdot 
(\bar N {\bf U})\right).
\eea
We see therefore that the momentum law is equivalent to  phonon-number conservation 
combined with the ray tracing equation
(\ref{EQ:kdot}).

\section{Discussion}

The possibilty of interpreting   the time average  of our momentum conservation
law in terms of quantum quasi-particles should warn us that we are dealing with  
pseudomomentum and not with newtonian momentum\cite{mcintyre}.
Nonetheless the quantity   $\vev{\rho_{(1)}\v_{(1)}}= \bar N\hbar {\bf k}$
is reliably computed from
the linearized wave equation, and is {\it part\/} of the true momentum. 
It is simply not all of it.  Even in the absence of a mean flow
with its $\vev{\rho_{(2)}\v}$ contribution  we still have to contend
with $\rho_{(0)}\vev{\v_{(2)}}$, and this  can be important.  As
an example \cite{mcintyre}, consider a closed cylinder filled with
fluid. At one end of the cylinder a piston is driven so as to
generate plane sound waves which completely span the cross section
of  the tube. At the other end a second piston is driven at the same
frequency with its phase adjusted  so as to absorb the sound waves
without reflection. It easy to see that  an extra
pressure   equal to $\vev{W_r}$ is exerted on the ends of the tube over
and above whatever isotropic pressure acts on the ends and
sides equally. It is ``obvious'' that this is   the force per
unit area $\bar N\hbar {\bf k}c$ required to generate and absorb the
phonon beam ``momentum''.  Unfortunately for this simple  idea, it is
equally obvious that the time average center-of-mass velocity of the
fluid in the tube vanishes, so the true momentum density in the beam is
exactly zero.  The $\vev{\rho_{(1)}\v_{(1)}}$ contribution to the
momentum density is exactly cancelled by a  $\rho_{(0)}\vev{\v_2}$
counterflow.  This eulerian streaming is driven by the  fluid source
term for $\vev{\v_{(2)}}$ implicit in (\ref{EQ:2nd_continuity})
\cite{lighthill_streaming}. (In a lagrangian description the
particles merely oscillate back and forth with no secular drift).
The momentum {\it flux\/} however is exactly the same {\it as if\/}
(the italics are from \cite{mcintyre}) there was no medium and the phonons were
particles possessing momentum $\hbar {\bf k}$. This is frequently
true: the flux of pseudomomentum is often equal to the flux of true
momentum to $O(a^2)$ accuracy.  Pseudomomentum flux can therefore be
used to compute forces.  On the other hand the  density of true
momentum in the fluid and the density of pseudomomentum are usually
unrelated\footnote{This does not mean that the attribution of
momentum to a phonon in the two-fluid model for  a superfluid is
incorrect. In  superfluid hydrodynamics the $\rho_{(0)}\vev{\v_2}$ counterflow is accounted
for separately from the $\vev{\rho_{(1)}\v_{(1)}}=\bar Nh\bar {\bf
k}$  normal-component mass flux. The counterflow is included in the
supercurrent  needed  to enforce $\nabla\cdot(\rho_n\v_n+
\rho_s\v_s)=0$.}.

It should be said that the $\rho_{(0)}\vev{\v_2}$ counterflow will
not always cancel the $\rho_{(1)}\vev{\v_1}$ wave pseudomomentum\cite{yih}. The
$\vev{\v_2}$ flow depends  the geometry. It is found from the source
equation (\ref{EQ:2nd_continuity}) and from the force the sound field
applies to the fluid.  The latter will be small when there is no
dissipation, as is the case in a superfluid, and for an isolated
sound beam source in an infinite medium $\vev{\v_2}$ will consist of
a  flow directed radially inwards towards the transducer  of
sufficient magnitude to supply the mass flowing out  along the sound
beam\cite{lighthill_streaming}. In the presence of dissipation the
force becomes important, leading to  acoustic streaming.

Consider our closed cylinder further. From   (\ref{EQ:true_rho2}) we
see that in a system with fixed $\vev{P}$, and in the presence of the sound wave, 
the mean density of the
fluid will be reduced by   
\be
\vev{\rho_{(2)}}= -\frac{\vev{W_r}}{c^2} \left.\frac{d\ln c}{d\ln \rho}\right|_{\rho_{(0)}}.
\ee
Since our cylinder has fixed volume, this density reduction cannot take place. Instead
it 
is opposed 
by a  pressure on the cylinder wall 
 \be
\Delta P=  \vev{W_r}\left.\frac{d\ln c}{d\ln
\rho}\right|_{\rho_{(0)}},
\ee
that must be added  to the   isotropic pressure in the absence of the sound
wave. The complete radiation stress tensor is therefore
\be
\vev{\Sigma_{ij}} = 
\vev{W_r}\left(\frac {k_ik_j}{k^2}+\delta_{ij}\frac{d\ln c}{d\ln
\rho}\right).
\ee
This result goes back to Brillouin\cite{brillouin}. 
The true radiation stress therefore differs from the
pseudomomentum flux tensor in its isotropic part. Forces computed from
pseudomomentum flux will therefore be incorrect if this pressure gradient is important. 
Usually it is not. See \cite{post} for examples.

\section{Acknowledgements}

This work was supported by grant NSF-DMR-98-17941. I would like to
thank Edouard Sonin,  David Thouless, and Ping Ao for many
discussions, and  Stefan Llewellyn-Smith for useful e-mail.

\appendix
\section{Geodesics and Hamiltonian Flows}

In this appendix we show that the null geodesics of the Unruh metric
coincide with conventional hamiltonian optics ray tracing.
The usual ray tracing equations are derived from $\omega({\bf k},{\bf
x})$ as
\be 
\dot {\bf x} = \frac {\partial \omega}{\partial {\bf k}}, 
\qquad \dot {\bf k} = -\frac {\partial \omega}{\partial {\bf x}}.
\ee
In our case  $ 
\omega({\bf k},{\bf x}) = c|k|+ {\bf v}\cdot {\bf k}
$.
Thus 
\be
\frac{dx^i}{dt}= v_{i} + c\frac { k_i}{|k|},\qquad
\frac{dk_i}{dt }= -\frac{\partial v_{j}}{\partial x^i} k_j.
\ee

We begin by noting that geodesics with an affine parameter $\tau$ are stationary paths
for the lagrangian   
\be
L= \frac 12 g_{\mu\nu}\frac{dx^\mu}{d\tau} \frac{dx^\nu}{d\tau}.
\ee
To make connection with the  ray tracing formalism  we consider  
the corresponding hamiltonian
\be
H =  \frac 12 g^{\mu\nu}p_\mu p_\nu,
\ee
and write down Hamilton's equations  with $\tau$ playing the role of time  
\bea
\frac{dx^\mu}{d\tau}&=& \frac{\partial H}{\partial p_\mu}=
g^{\mu\nu}p_\nu\cr
 \frac{dp_\mu}{d\tau}&=& - \frac{\partial H}{\partial x^\mu}=-\frac 12
\frac{\partial g^{\alpha\beta}}{\partial x^\mu} p_\alpha p_\beta.
\eea
Combining  these gives
\be
\frac {d^2 x^\mu}{d\tau^2} = \frac {\partial g^{\mu\beta}}{\partial x^\alpha}\frac{dx^\alpha}{d\tau}
p_\mu+ g^{\mu\nu} \left( -\frac 12 \frac {\partial g^{\alpha\beta}}{\partial
x^\nu}\right) p_\alpha p_\beta.
\ee
Now for matrices ${\bf g}$ we have
\be
d{\bf g}^{-1} =- {\bf g}^{-1} (d{\bf g}) {\bf g}^{-1},
\ee
so with $({\bf g})_{\alpha\beta}=g_{\alpha\beta}$ and $({\bf
g}^{-1})_{\alpha\beta}=g^{\alpha\beta}$ we can write
\be
\frac {d^2 x^\mu}{d\tau^2} +\frac 12 g^{\mu\nu}\left( \frac{\partial
g_{\nu\alpha}}{\partial x^\beta} + \frac{\partial
g_{\nu\beta}}{\partial x^\alpha} - \frac{\partial
g_{\alpha\beta}}{\partial
x^\nu}\right)\frac{dx^\alpha}{d\tau}\frac{dx^\beta}{d\tau}=0,
\ee  
which is the geodesic equation.

We now examine these equations for  the particular case of the Unruh metric. 
We define a $4$-vector $p_\mu=( \omega, -k_i)$ so that
$p_\mu x^\mu = \omega t - {\bf k}\cdot {\bf x}$. Then 
\be 
H =  \frac 12 g^{\mu\nu}p_\mu p_\nu  
= \frac 12 \left( (\omega - \v\cdot {\bf k} )^2 -c^2 |k|^2\right).
\ee
Hamilton's equations become
\be
\frac{dx^0}{d\tau} = \frac{dt}{d\tau}= \frac{\partial H}{\partial \omega}
= \omega  - \v\cdot {\bf k},
\ee
and
\be   
\frac{dx^i}{d\tau}= - \frac{\partial H}{\partial k_i}= 
v_{i} (\omega - \v\cdot {\bf k}) + c^2  k_i.
\ee
For null geodesics  $(\omega - \v\cdot {\bf k})^2 -c^2 |k|^2=0$, or $(\omega - \v\cdot
{\bf k})=c|k|$. Thus
\be
\frac{dx^i}{dt} = v_{i} + \frac {c^2 k_i}{(\omega - \v\cdot {\bf k})},
\ee
or
\be
\frac{dx^i}{dt}= v_{i} + c\frac { k_i}{|k|},
\ee
which is the group velocity  equation.
We also find
\be
\frac{d\omega}{d\tau}= -\frac{\partial H}{\partial t}=0
\ee
if the flow is steady, and
\be
-\frac{dk_i}{d\tau}=  -\frac{\partial H}{\partial x^i}=(\omega - \v\cdot {\bf k})
\frac{\partial v_{j}}{\partial x^i} k_j,
\ee
which is equivalent to the momentum evolution equation
\be
\frac{dk_i}{dt }= -\frac{\partial v_{j}}{\partial x^i} k_j.
\label{EQ:kdot}
\ee


\eject
\end{document}